\documentclass[12pt]{article}
\usepackage{graphicx}
\newsavebox{\sboxpubnumber}
\newsavebox{\sboxpubdate}

\newcommand{\pubnumber}[1]{\begin{lrbox}{\sboxpubnumber}{\begin{tabular}{l}
#1 \\
				 \usebox{\sboxpubdate}
				 \end{tabular}}
                           \end{lrbox}
                           \pubblock}
\newcommand{\Title}[1]{\begin{center} {\Large #1 } \end{center}}
\newcommand{\Author}[1]{\begin{center}{ \sc #1} \end{center}}
\newcommand{\Address}[1]{\begin{center}{ \it #1} \end{center}}

\newcommand{\pubblock}{\rightline{
			\usebox{\sboxpubnumber}}}

\newenvironment{Presented}{\begin{quotation} \begin{center}
             PRESENTED AT\end{center}\bigskip
      \begin{center}\begin{large}}{\end{large}\end{center}
      \end{quotation}}
\newcommand{\Acknowledgements}{\bigskip  \bigskip \begin{center} \begin{large}
             \bf ACKNOWLEDGEMENTS \end{large}\end{center}}

\begin{document}
\begin{titlepage}
\pubnumber{UNIL-IPT 01-17 \\ YYY-YYYYYY} 
\vfill
\Title{Graviphoton and graviscalars delocalization\\ in brane world
scenarios}
\vfill
\Author{Massimo Giovannini\footnote{Electronic address: 
Massimo.Giovannini@ipt.unil.ch}}
\Address{Institute of Theoretical Physics, University of Lausanne, \\
BSP CH-1015, Dorigny, Switzerland}
\vfill
\begin{abstract}
A manifestly gauge-invariant theory of gravitational fluctuations 
of brane-world scenarios is discussed. Without resorting 
to any specific gauge choice, a general method is presented 
in order to disentangle the fluctuations of the brane energy-momentum 
from the fluctuations of the metric.
As an application of the formalism, 
the localization  of metric fluctuations on scalar branes 
breaking spontaneously five-dimensional Poincar\'e invariance is addressed. 
Only assuming that the four-dimensional Planck mass is finite and that the
geometry is regular, it is demonstrated that the vector and scalar 
fluctuations of the metric are not localized on the brane. 
\end{abstract}
\begin{Presented}
    COSMO-01\\
    Rovaniemi, Finland, \\
    August 29 -- September 4, 2001
\end{Presented}
\vfill
\end{titlepage}
\def\thefootnote{\fnsymbol{footnote}}
\setcounter{footnote}{0}

\renewcommand{\theequation}{1.\arabic{equation}}
\setcounter{equation}{0}
\section{Introduction}
Fields of various spin can be localized \cite{m1} on higher dimensional 
topological defects \cite{m2,ak,vis}. A field is localized if 
it exhibits a normalizable zero mode with respect 
to the bulk coordinates parameterizing the geometry of 
the defect in the extra-dimensional space. 
Moreover, if the four-dimensional Planck mass is finite, the 
{\em tensor} modes of the geometry itself can be localized 
on a higher dimensional topological defect living, respectively, 
in six, seven or eight dimensions \cite{misproc}. An 
example of this phenomenon is provided 
by five-dimensional AdS space with a brane source \cite{rs,rs2}.

The question which will be addressed in this paper
 is the fate of the other modes of the geometry 
itself, namely the scalar and vector fluctuations. In order 
to make the discussion more concrete let us concentrate on the case 
of a $ D = d + 2$-dimensional geometry characterized only 
by one bulk coordinate $w$ and whose line element can be written as 
\begin{equation}
ds^2 = \overline{G}_{A B} dx^A dx^B =
 a^2(w)[ dt^2 - d x_{1}^2 -...- d x_{d}^2 - dw^2],
\label{me}
\end{equation}
where $a(w)$ is the warp factor and $\overline{G}_{A B}$ denotes the 
background metric. For a review on the 
various implications of warped  geometries see also \cite{rub}.

The analysis of gravitational fluctuations in brane-world
models may be rather complicated. Given a physical 
brane configuration characterized by a thickness and a 
profile, the fluctuations of the geometry will be
necessarily entangled with the fluctuations of the 
energy-momentum tensor of the brane. Furthermore, the 
fluctuations of the metric (and of the brane energy-momentum tensor) 
depend upon the coordinate system. Both problems 
may a have a crucial impact on the analysis of the fluctuations. 
If 
the normal modes of the system are not properly analyzed, their localization 
cannot be discussed. 

Since the meaning of the 
fluctuations may change  in different coordinate 
systems, spurious gauge modes could make the whole 
analysis unreliable. 
Similar caveats should be borne in mind in the analysis of metric 
fluctuations on a given cosmological background in 
four-dimensions. Here the situation is similar but also 
crucially different: because of the five-dimensional 
nature of the geometry, the metric perturbations have to be 
classified according to the symmetry of the problem. If 
five-dimensional Poincar\'e invariance is broken (either 
spontaneously or explicitly), the various 
modes of the geometry should be classified according to four-dimensional 
Poincar\'e transformations. In cosmology 
fluctuations are classified according to the group of rotations 
in three dimensions.

In order to discuss the metric fluctuations of scalar-tensor 
actions, the {\em Bardeen formalism} \cite{bar} has been very useful in 
four-dimensional backgrounds. The main idea is to parameterize 
the metric fluctuations by defining a suitable set of gauge-invariant 
variables which do not change for infinitesimal coordinate 
transformations. The Bardeen formalism is also rather effective in order 
to identify the coordinate systems where the gauge-invariant 
variables take a simple form.  In the following it will be shown that 
a non-trivial generalization of the formalism 
is allowed by requiring that the geometry has more than four 
 dimensions and by classifying the fluctuations with respect 
to four-dimensional  Poincar\'e transformations \cite{n1}.

Having generalized the Bardeen 
formalism to the case of five-dimensional warped geometries, 
the localization properties of metric 
fluctuations can be investigated in a fully gauge-invariant
manner. If the zero mode of a given fluctuation is 
not normalizable, then it will not be localized on the brane and 
it will not affect the four-dimensional physics. Since 
all the equations for the fluctuations will be reduced 
to second order (partial) dfferential equations, the tower 
of mass eigenstates can be analyzed using known methods 
borrowed from supersymmetric quantum mechanics \cite{susqm}.

Explicit physical (thick) brane solutions have  been derived 
in different numbers of transverse dimensions. In five dimensions 
physical branes can be obtained in terms  of a scalar domain-wall whose 
 scalar field depends upon the bulk radius \cite{kt1,kt2,gremm1,gremm2}. 
In \cite{free,free2} the stability of scalar domain walls (inspired by gauged 
supergravity theories) has been analyzed. 
In more than five dimensions, physical brane solutions including  also 
 background gauge fields have been recently discussed \cite{gms,gm}. 
The formalism presented in this paper can be generalized to the case 
where the transverse space contains more than one bulk coordinate. 
This formalism has been also applied in the 
discussion of the radion wave-function in connection 
with the self-tuning problem \cite{kim} and it has been 
recently generalized to the case where 
the Einstein-Hilbert 
action is supplemented by quadratic 
curvature corrections parametrized in the Euler-Gauss-Bonnet (EGB) 
form \cite{n3}. Here a self-contained derivation, heavily based 
on gauge-invariance, will be presented.

The plan of this paper is the following. In Section II 
 gauge-invariance will be exploited in order to obtain a set of 
evolution equations which are independent on the specific 
coordinate system. In Section III the general formalism 
will be applied to the case of thick brane configurations. Section IV contains 
our concluding remarks. 

\renewcommand{\theequation}{2.\arabic{equation}}
\setcounter{equation}{0}
\section{Exploiting gauge-invariance} 
The equations describing the brane configuration can be written, in 
general terms, as
\begin{equation}
R_{A}^{B} = 8 \pi G_{D} ~\tau_{A}^{B},
\label{ce}
\end{equation}
where $R_{A}^{B}$ is the Ricci tensor and 
\begin{equation}
\tau_{A}^{B} = 8 \pi G_{D}~(T_{A}^{B} - 
\frac{T}{d}\delta_{A}^{B}), \,\,\,\, T= T_{A}^{A},
\end{equation}
is constructed from the energy-momentum tensor of the brane $T_{A}^{B}$.

For the background metric of  Eq. (\ref{me}) describing a warped 
space-time the Ricci tensors are \footnote{The Greek  indices run over the 
$d+1$-dimensional  sub-space whereas the Latin indices run over the full 
$D$-dimensional space-time.}  
\begin{eqnarray}
&& \overline{R}_{\mu}^{\nu} =  \frac{1}{a^2} ( {\cal H}' + d {\cal H}^2) 
\delta_{\mu}^{\nu},
\nonumber\\
&&\overline{R}_{w}^{w} = \frac{(d +1)}{a^2} {\cal H}' ,
\label{ricci}
\end{eqnarray}
where ${\cal H} = a'/a$ and the prime denotes derivation with 
respect to $w$.
Then the background equations can be written as
\begin{equation}
\overline{R}_{A}^{B} = 8 \pi G_{D} \overline{\tau}_{A}^{B} ,
\label{back}
\end{equation}
 where $\overline{\tau}_{A}^{B}$ is the background energy-momentum 
tensor. In spite of the specific 
form of the action describing 
the domain-wall solution, its energy energy-momentum tensor should satisfy the 
following symmetry properties
\begin{eqnarray}
&&\overline{\tau}_{\mu}^{w} = \overline{\tau}_{w}^{\mu} =0,
\nonumber\\
&&\overline{\tau}_{\mu}^{\nu} \propto \delta_{\mu}^{\nu}, 
\nonumber\\
&&\overline{\tau}_{w}^{w} \propto \delta_{w}^{w}. 
\label{backtens}
\end{eqnarray}
which are obtained from Eqs. (\ref{back}) by requiring that the 
form of $\tau_{A}^{B}$ is compatible with the specific form of the Ricci 
tensor given in Eq. (\ref{ricci}).

To first order in the fluctuations of the metric
\begin{equation}
G_{A}^{B}(x^{\mu}, w) = \overline{G}_{A}^{B}(w) + \delta G_{A}^{B}(x^{\mu}, w),
\end{equation}
the fluctuations of the Ricci tensors and of the $\tau_{A}^{B}$ can be 
written as 
\begin{eqnarray}
&&
R_{A}^{B} (x^{\mu}, w) = \overline{R}_{A}^{B}(w) +\delta R_{A}^{B}(x^{\mu}, w),
\label{ricpert}
\nonumber\\
&& 
\tau_{A}^{B} (x^{\mu}, w) = \overline{\tau}_{A}^{B}(w) 
+\delta \tau_{A}^{B}(x^{\mu}, w),
\end{eqnarray}
where $ \delta$ denotes all the terms linear in the metric and matter 
fluctuations.
The equations of motion for small metric fluctuations linearized 
around the background are
\begin{equation}
\delta R_{A}^{B} = 8 \pi G_{D} \delta \tau_{A}^{B}. 
\label{perteq}
\end{equation}

In general the fluctuations of the Ricci tensor and of the energy-momentum 
tensor will have scalar, vector and tensor modes which 
can be classified according to the $(d+1)$-dimensional Poincar\'e invariance 
of the metric (\ref{me}).
Without assuming any specific gauge the fluctuations of the 
Ricci tensor can be obtained, after a tedious calculation involving 
the repeated use of Palatini identities (see the Appendix).
In particular it can be obtained that 
\begin{eqnarray}
\delta R_{w}^{w} &=& \frac{(d + 1)}{a^2}\biggl\{ \psi'' + {\cal H} (\xi' 
+ \psi' ) + 2 {\cal H}' \xi 
\nonumber\\
&-& \partial_{\alpha} \partial^{\alpha}[ ( C - E')' 
+ {\cal H} ( C - E') - \xi]\biggr\} , 
\label{p1}\\
\delta R_{\mu}^{w} &=& \frac{1}{a^2} \biggl\{ d \partial_{\mu} ( {\cal H} \xi 
+ \psi') 
+ \frac{1}{2} \partial_{\alpha}\partial^{\alpha} ( D_{\mu} - 
f_{\mu}') \biggr\},
\label{p2}\\
\delta R_{\mu}^{\nu} &=& \frac{1}{a^2} \biggl\{ {h_{\mu}^{\nu}}'' + 
d {\cal H} {h_{\mu}^{\nu}}'  - \partial_{\alpha}\partial^{\alpha} 
h_{\mu}^{\nu} 
\nonumber\\
&+&\delta_{\mu}^{\nu} \bigl[ \psi'' + ( 2 d + 1) {\cal H} \psi' 
- \partial_{\alpha} \partial^{\alpha} \psi 
\nonumber\\
&+& {\cal H} \xi' + 2 ({\cal H}' + d {\cal H}^2 ) \xi - 
{\cal H} \partial_{\alpha}\partial^{\alpha} ( C - E') \bigr]  
\nonumber\\
&+&  
\partial_{\mu} \partial^{\nu}[ (E' - C)' + d {\cal H} (E' - C) + \xi
- (d -1) \psi]
\nonumber\\
&+& [ (\partial_{(\mu}f^{\nu)})'' + d {\cal H}  (\partial_{(\mu}f^{\nu)})']
- [ (\partial_{(\mu}D^{\nu)})' + d {\cal H}  (\partial_{(\mu}D^{\nu)})]
\biggr\}. 
\label{p3}
\end{eqnarray}
The various functions appearing in Eqs. (\ref{p1})--(\ref{p2}) come from the 
perturbed form of the metric 
\begin{equation}
\delta G_{A B}=a^2(w) \left(\matrix{2 h_{\mu\nu} 
+(\partial_{\mu} f_{\nu} +\partial_{\nu} f_{\mu}) 
+ 2\eta_{\mu \nu} \psi
+ 2 \partial_{\mu}\partial_{\nu} E
& D_{\mu} + \partial_{\mu} C &\cr
D_{\mu} + \partial_{\mu} C  & 2 \xi &\cr}\right).
\label{pm}
\end{equation}
In Eq. (\ref{pm}) $h_{\mu\nu}$ is a divergence-less and trace-less 
[i.e.$\partial_{\mu} h^{\mu}_{\nu} =0$, $h_{\mu}^{\mu} =0$]
rank-two tensor in the $(d+1)$-dimensional Poincar\'e invariant 
space-time. The vectors $f_{\mu}$ and $D_{\mu}$ are both divergence-less
[i.e. $\partial_{\mu} D^{\mu} =\partial_{\nu} f^{\nu} =0$]. 
The other functions are 
scalars under Poincar\'e transformations. 
The number of independent functions parameterizing the fluctuations 
of the metric is then $(d+2)(d+3)/2$.

The scalar, vector and tensor fluctuations of the geometry defined in 
Eq. (\ref{pm}) are not invariant under infinitesimal coordinate 
transformations of the type 
\begin{equation}
x^{A} \rightarrow \tilde{x}^{A} = x^{A} + \epsilon^{A},
\label{trans}
\end{equation}
where $\epsilon^{A}$ stands for the $(d+2)$ functions 
parameterizing the gauge transformation.
Similarly to what happens for the metric 
fluctuations, the perturbations of the Ricci tensor change 
for infinitesimal coordinate transformations. 
Hence, the logic will now be the following.
 From the  transformation properties of 
the metric fluctuations under the gauge shift of Eq. (\ref{trans}) 
gauge-invariant functions will be defined. This definition 
should be simple (i.e. linear in the fluctuations) and moreover 
it should allow to write the fluctuations 
of the Ricci tensor in terms of a gauge-invariant part supplemented 
by a gauge-dependent part. It will be shown, in general, that 
the gauge-dependent part of the Ricci fluctuation is identically 
  compensated by the gauge-dependent part of the fluctuation of the 
brane energy-momentum tensor. This observation allows to 
write, in general terms, a gauge-invariant form of the 
evolution equations of the 
fluctuations. The obtained system, will be, by construction, 
independent on the specific coordinate system.

Under the transformation of Eq. (\ref{trans}) the 
fluctuations of the metric transform according to the 
usual expression involving the Lie derivative in the direction 
of the vector $\epsilon^{A}$
\begin{equation}
\delta \tilde{G}_{A B} = \delta G_{AB} - \nabla_{A} \epsilon_{B} - \nabla_{B}
\epsilon_{A},
\label{liederiv}
\end{equation}
where $\epsilon_A = a^2(w) (\epsilon_{\mu}, -\epsilon_{w})$. We can always 
use $(d+1)$-dimensional Poincar\'e invariance in order to 
classify the transformation properties of the gauge functions. In pratical 
terms, 
\begin{equation}
\epsilon_{\mu} = \partial_{\mu} \epsilon + \zeta_{\mu},
\end{equation}
with $\partial_{\mu} \zeta^{\mu} =0$. Hence, while $\zeta_{\mu}$ 
transforms as a pure (i.e. divergence-less vector), $\epsilon $ and 
$\epsilon_{w}$ are scalars.

The tensor mode of the metric, i.e.
$h_{\mu\nu}$ are automatically invariant under the transformation 
(\ref{trans}).  The vectors and the scalars are not gauge-invariant and this 
is the source of the lack of gauge-invariance of Eqs. (\ref{p1})--(\ref{p3}).
Since there are two scalar gauge functions  [i.e. 
$\epsilon$ and $\epsilon_{w}$],
and one vector gauge function [i.e. $\zeta_{\mu}$], two
gauge-invariant scalar functions and one gauge-invariant vector function
can be defined. The gauge-invariant scalars are
\begin{eqnarray}
&&\Psi = \psi - {\cal H}  ( E' - C), 
\label{giscal0}\\
&& \Xi = \xi - \frac{1}{a} [ a( C - E')]'.
\label{giscal}
\end{eqnarray}
The gauge-invariant vector is
\begin{equation}
V_{\mu} = D_{\mu} - f_{\mu}'.
\label{givec}
\end{equation}
By noticing that, under (\ref{trans}) the scalar fluctuations of 
(\ref{pm}) transform as 
\begin{eqnarray}
&&E\to\tilde{E} = E - \epsilon,
\label{El}\\
&&\psi\to \tilde{\psi} = \psi - {\cal H} \epsilon_{w},
\label{psil}\\
&& C\to\tilde{C} = C - \epsilon' + \epsilon_{w},
\label{Cl}\\
&& \xi\to \tilde{\xi} = \xi + {\cal H} \epsilon_{w} + \epsilon_{w}',
\label{xil}
\end{eqnarray}
and the vector fluctuations transform as 
\begin{eqnarray}
&& f_{\mu}\to \tilde{f}_{\mu} = f_{\mu} - \zeta_{\mu},
\label{fl}\\
&& D_{\mu}\to \tilde{D}_{\mu} = D_{\mu} - \zeta_{\mu}',
\label{zeta}
\end{eqnarray}
the gauge-invariance of Eqs. (\ref{giscal0})--(\ref{giscal})  
and (\ref{givec}) can be explicitly shown.

Using Eqs. (\ref{giscal0})--(\ref{givec}) into Eqs. (\ref{p1})--(\ref{p3}) 
the fluctuations of the Ricci tensor can be written in a fully gauge-invariant 
manner
\begin{eqnarray}
&&\delta R_{w}^{w} = \delta^{{\rm (gi)}} R_{w}^{w} - \bigl[\overline{R}_{w}^{w}
\bigr]'( C - E'), 
\label{g1}\\
&& \delta R_{\mu}^{w} 
= \delta^{{\rm (gi)}} R_{\mu}^{w} - \overline{R}_{w}^{w} 
\partial_{\mu}( C - E') + \overline{R}_{\mu}^{\nu} \partial_{\nu} ( C - E'),
\label{g2}\\
&&  \delta R_{\mu}^{\nu} 
= \delta^{{\rm (gi)}} R_{\mu}^{\nu} - [\overline{R}_{\mu}^{\nu}]' 
( C - E'),
\label{g3}
\end{eqnarray}
where $\delta^{{\rm (gi)}}$ denotes a variation which 
preserves gauge-invariance and where
\begin{eqnarray}
\delta^{{\rm (gi)}} R_{w}^{w}   &=& \frac{1}{a^2} \biggl\{ (d + 1)
[\Psi'' + {\cal H} ( \Psi' + \Xi') + 2 {\cal H}' \Xi] + 
\partial_{\alpha}\partial^{\alpha} \Xi\biggr\}
\label{gir1}\\
\delta^{{\rm (gi)}} R_{\mu}^{w} &=& \frac{1}{a^2}\biggl\{ d \partial_{\mu}[
{\cal H} \Xi + \Psi'] + \frac{1}{2} \partial_{\alpha}\partial^{\alpha} V_{\mu}
\biggr\},
\label{gir2}\\
 \delta^{{\rm (gi)}} R_{\mu}^{\nu} &=& \frac{1}{a^2} \biggl\{  
\partial_{\mu} \partial^{\nu} [ \Xi 
- ( d - 1) \Psi]
\nonumber\\
&+&\delta_{\mu}^{\nu} [ \Psi'' + ( 2 d + 1) {\cal H} \Psi' - 
\partial_{\alpha} \partial^{\alpha} \Psi + {\cal H} \Xi' + 
2 ( {\cal H}' + d {\cal H}^2) \Xi] 
\nonumber\\
&-& [ (\partial_{(\mu}V^{\nu)})' + d {\cal H} 
(\partial_{(\mu} V^{\nu)})] +{h_{\mu}^{\nu}}''
+ d {\cal H} {h_{\mu}^{\nu}}' - \partial_{\alpha} \partial^{\alpha} 
h_{\mu}^{\nu}  \biggr\}.
\label{gir3}
\end{eqnarray}
Bearing in mind the symmetry properties of the background energy-momentum 
tensor we can also write the fluctuations of $\tau_{A}^{B}$ in 
gauge-invariant terms:
\begin{eqnarray}
\delta \tau_{w}^{w} &=& \delta^{({\rm gi})} \tau_{w}^{w} -
 \bigl[\overline{\tau}_{w}^{w}\bigr]'( C - E'), 
\label{t1}\\
\delta \tau_{\mu}^{w} 
&=& \delta^{{\rm (gi)}} \tau_{\mu}^{w} - \overline{\tau}_{w}^{w} 
\partial_{\mu}( C - E') + \overline{\tau}_{\mu}^{\nu} \partial_{\nu} ( C - E'),
\label{t2}\\
\delta \tau_{\mu}^{\nu} 
&=& \delta^{{\rm (gi)}} R_{\mu}^{\nu} - [\overline{\tau}_{\mu}^{\nu}]' 
( C - E').
\label{t3}
\end{eqnarray}
Inserting now Eqs. (\ref{g1})--(\ref{g3}) and Eqs. (\ref{t1})--(\ref{t3}) 
into Eqs. (\ref{perteq}) we obtain, after the use of Eqs. 
(\ref{back}), the gauge-invariant form of the perturbed equations, namely
\begin{eqnarray}
&&(d + 1)[\Psi'' + {\cal H} ( \Psi' + \Xi') + 2 {\cal H}' \Xi] + 
\partial_{\alpha}\partial^{\alpha} \Xi = 8 \pi G_{D} a^2 \delta^{({\rm gi})}
\tau_{w}^{w}
\label{eq1}\\
&&  d \partial_{\mu}[
{\cal H} \Xi + \Psi'] + \frac{1}{2} \partial_{\alpha}\partial^{\alpha} V_{\mu}
=8 \pi G_{D} a^2 \delta^{({\rm gi})}
\tau_{\mu}^{w}
\label{eq2}\\
&& {h_{\mu}^{\nu}}''
+ d {\cal H} {h_{\mu}^{\nu}}' - \partial_{\alpha} \partial^{\alpha} 
h_{\mu}^{\nu}  
\nonumber\\
&+& \delta_{\mu}^{\nu} [ \Psi'' + ( 2 d + 1) {\cal H} \Psi' - 
\partial_{\alpha} \partial^{\alpha} \Psi + {\cal H} \Xi' + 
2 ( {\cal H}' + d {\cal H}^2) \Xi] 
\nonumber\\
&+& \partial_{\mu} \partial^{\nu} [ \Xi 
- ( d - 1) \Psi] - [ (\partial_{(\mu}V^{\nu)})' + d {\cal H} 
(\partial_{(\mu} V^{\nu)})]=8 \pi G_{D} a^2 \delta^{({\rm gi})}
\tau_{\mu}^{\nu}.
\label{eq3}
\end{eqnarray}
The gauge invariant variation of the brane energy-momentum 
tensor can be obtained once the brane action is specified and specific 
examples will be provided in the following Section. In spite of this
it should be noticed that the derivation presented in this 
paper is completely independent on the specific form of the 
brane energy-momentum tensor. It is only required that the 
generic brane configuration is a solution of the 
five-dimensional extension of the Einstein-Hilbert theory.

\renewcommand{\theequation}{3.\arabic{equation}}
\setcounter{equation}{0}
\section{Localization of the modes of the geometry} 

The formalism discussed in the previous 
section is based on the construction of a gauge-invariant fluctuation
of the evolution equations describing the 
brane configuration.
The following five-dimensional action \footnote{Notice that 
 $\kappa = 8\pi G_{5} = 8\pi/M_{5}^3$. Natural gravitational 
units will be often employed by setting $2 \kappa =1$.}. 
\begin{equation}
S= \int d^{5}x \sqrt{|G|}\biggl[- \frac{R}{2\kappa} + \frac{1}{2} G^{A B} 
\partial_{A} \varphi \partial_{B} \varphi - V(\varphi)\biggr],
\label{ac}
\end{equation}
can be used in order to describe the breaking of five-dimensional Poincar\'e 
symmetry. Consider a potential which is invariant under the 
$\varphi \rightarrow -\varphi$ symmetry. Then, non-singular domain-wall 
solutions can be obtained, for various potentials.
For instance solutions of the type 
\begin{eqnarray}
&& a(w) = \frac{1}{\sqrt{b^2 w^2 + 1}},
\label{s1}\\
&& \varphi = \varphi(w) = \sqrt{2 d} \arctan{b w},
\label{s2}
\end{eqnarray}
can be found for different classes of symmetry-breaking 
 potentials \cite{gremm1,kt1,free}.  Solutions like Eqs. 
(\ref{s1})--(\ref{s2}) represent a smooth  version of the Randall-Sundrum
scenario \cite{rs,rs2}.
The assumptions of the present analysis will now be listed. 

(i) The five-dimensional geometry is regular (in a technical sense)
for any value of the bulk coordinate $w$. This implies that singularities 
in the curvature invariants are absent.

(ii) D-dimensional Poincar\'e invariance is broken 
through a smooth five dimensional domain-wall solution generated 
by a potential $V(\varphi)$ which is invariant under 
$\varphi \rightarrow - \varphi$. The warp factor $a(w)$ will 
then be assumed symmetric for $w\rightarrow - w$.

(iii) Four-dimensional Planck mass is finite because the following integral
converges
\begin{equation}
M^2_{P} \simeq M^d \int_{-\infty}^{\infty} dw a^d(w).
\label{pl}
\end{equation}

(iv) Five-dimensional gravity is described according to Eq. (\ref{ac}) 
and, consequently, the equations of motion for the warped background 
generated by the smooth wall are, in natural gravitational units, 
\begin{eqnarray}
&&{\varphi'}^2 = 2 d ( {\cal H}^2 - {\cal H}'), 
\label{c1}\\
&& Va^2 = - d ( d {\cal H}^2 + {\cal H}').
\label{c2}
\end{eqnarray}
where the prime denotes derivation with respect to $w$ and ${\cal H} = a'/a$.

Under these assumptions it will be shown that the gauge-invariant 
fluctuations corresponding to scalar and vector modes of the geometry 
are not localized on the wall. On the contrary, tensor modes of the geometry 
will be shown to be localized. This program will be achieved in two steps. 
In the first step decoupled equations for the gauge-invariant 
variables describing the fluctuations of the geometry will be obtained 
for general brane backgrounds
without assuming any specific solution.
In the second step the normalizability of the zero modes will be  
addressed using only the assumptions (i)--(iv).

Recalling that, in the case of the action (\ref{ac}) 
\begin{equation}
\tau_{A}^{B} = \partial_{A} \varphi \partial^{B} \varphi - 2\frac{V}{d} 
\delta_{A}^{B}
\end{equation}
the formalism discussed in the previous Section can be applied.
Consider now, for sake of simplicity, the case $d=3$ (the same 
can be however discussed for a generic $d$).
In this case, following the notations of the previous Section we get :
\begin{eqnarray}
&&\overline{\tau}_{\mu}^{\nu} = - \frac{2}{3} V \delta _{\mu}^{\nu},
\nonumber\\
&& \overline{\tau}_{w}^{w} = - \frac{{\varphi'}^2 }{a^2 } - \frac{2}{3} V,
\nonumber\\
&& \overline{\tau}_{\mu}^{w} =0.
\label{te1}
\end{eqnarray} 
The gauge-invariant fluctuation  of $\tau_{A}^{B}$ will instead be:
\begin{eqnarray}
&&  \delta^{(gi)} \tau_{\mu}^{\nu} = - \frac{2}{3} \frac{\partial V}{\partial\varphi} X \delta_{\mu}^{\nu},
\nonumber\\
&& \delta^{(gi)} \tau_{w}^{w} = - \frac{2}{a^2} \Xi - \frac{2}{a^2} \varphi' X' - \frac{2}{3} \frac{\partial V}{\partial\varphi} X,
\nonumber\\
&& \delta^{(gi)} \tau_{\mu}^{w} = - \frac{\varphi'}{a^2} \partial_{\mu} X,
\label{te2}
\end{eqnarray}
where 
\begin{equation}
X = \chi - \varphi' (E' -C).
\label{chigi}
\end{equation}
is the gauge-invariant fluctuation of the scalar field $\varphi$.
Hence, using Eqs. (\ref{t1}) and (\ref{t2}) into Eqs. (\ref{eq1})--(\ref{eq3}) 
we get
\begin{eqnarray}
&&\Psi'' + 7 {\cal H} \,\Psi' + {\cal H}\, \Xi' + 
2 ( {\cal H}' + 3 {\cal H}^2)\,\Xi + \frac{1}{3} 
\frac{\partial V}{\partial\varphi} a^2 X - \partial_{\alpha}
\partial^{\alpha} \Psi =0,
\label{ps}\\
&&- \partial_{\alpha}\partial^{\alpha}\,\Xi - 4\, [ \Psi''+ {\cal H}\,\Psi'] - 
4 {\cal H}\, \Xi' - \varphi' X' - \frac{1}{3} 
\frac{\partial V}{\partial\varphi} a^2 X + \frac{2}{3} V a^2 \Xi =0.
\label{xi}
\end{eqnarray}
Eqs. (\ref{ps}) and (\ref{xi}) are subjected to the constraints 
\begin{eqnarray}
&&\partial_{\mu}\partial_{\nu}[ \Xi - 2 \Psi] =0,
\label{con1}\\
&& 6 {\cal H}\, \Xi + 6 \Psi' + X \,\varphi' =0.
\label{con3}
\end{eqnarray}
These equations should be 
supplemented by the gauge-invariant version of the 
 perturbed scalar field equation 
\begin{equation}
\delta G^{A B} \,\biggl( \partial_{A} \partial_{B} \varphi -
\overline{\Gamma}_{A B}^{C}\, \partial_{C} \varphi \biggr) 
+ \overline{G}^{A B} \,\biggl( \partial_{A} \partial_{B} \chi - 
\overline{\Gamma}_{A B}^{C} \partial_{C} \chi  - \delta 
\Gamma^{C}_{A B} \partial\varphi\biggr) 
+ \frac{\partial^2 V}{\partial \varphi^2}\chi=0,
\label{P2}
\end{equation}
where $ \delta \Gamma^{C}_{A B}$ are the fluctuations of the connections.
The explicit (gauge-invariant)  version of Eq. (\ref{P2}) is 
\begin{equation}
\partial_{\alpha} \partial^{\alpha} X - X'' - 3 {\cal H} X' + 
\frac{ \partial^2 V}{\partial\varphi^2} a^2 X - 
\varphi' [ 4 \Psi' + \Xi'] - 2 \Xi\,( \varphi'' + 3 {\cal H} \varphi') =0.
\label{ch}
\end{equation}

The evolution of the gauge-invariant vector variable (\ref{givec}) is 
\begin{equation}
\partial_{\alpha} \partial^{\alpha} {\cal V}_{\mu} =0,\,\,\,\,\,\,
{\cal V}_{\mu}' + 
\frac{3}{2} {\cal H} {\cal V}_{\mu}=0,
\label{calV}
\end{equation}
where ${\cal V}_{\mu} = a^{3/2} V_{\mu}$ is the canonical 
normal mode of the action (\ref{ac}) perturbed to second order in the 
amplitude of vector fluctuations of the metric. 

The equation for the 
tensors, as expected, decouples from the very beginning:
\begin{equation}
\mu_{\mu\nu}'' - \partial_{\alpha}\partial^{\alpha} \mu_{\mu\nu} 
- \frac{(a^{3/2})''}{a^{3/2}} \mu_{\mu\nu} =0.
\label{mu}
\end{equation}
where $\mu_{\mu\nu} =a^{3/2} h_{\mu\nu}$ is the canonical 
normal mode of the of the action (\ref{ac}) perturbed to second order 
in the amplitude of tensor fluctuations.

Using repeatedly the constraints of Eqs. (\ref{con1})--(\ref{con3}), 
together with the background relations (\ref{c1})--(\ref{c2}), 
the scalar system can be reduced to the following two equations \cite{n1,n2}
\begin{eqnarray}
&& \Phi'' - \partial_{\alpha}\partial^{\alpha} \Phi - 
z\biggl(\frac{1}{z}\biggr)'' \Phi =0,
\label{ph}\\
&& {\cal G}'' - \partial_{\alpha}\partial^{\alpha}{\cal G}
- \frac{z''}{z} {\cal G} =0,
\label{G}
\end{eqnarray}
where 
\begin{equation}
\Phi= \frac{a^{3/2}}{\varphi'} \Psi,\,\,\,\,\,\,\,{\cal G} = 
a^{3/2} X - z \Psi.
\label{def}
\end{equation}
The same equation satisfied by $\Psi$ is also satisfied by $\Xi$ by virtue 
of the constraint (\ref{con1}).
In Eq. (\ref{G}) and (\ref{def}) the background dependence appears in terms 
of the ``universal'' function $z(w)$ 
\begin{equation}
z(w) = \frac{a^{3/2} \varphi'}{{\cal H}}.
\label{z}
\end{equation}
Notice, incidentally, that ${\cal G}$ represent the canonical normal modes 
of the action.
If the  action (\ref{ac}) is perturbed to second order in the 
amplitude 
of the scalar fluctuations of the system, then its form, up to total 
derivatives, is
\begin{equation}
\delta^{(2)} S_{S} = \int d^4 x d w \frac{1}{2}\biggl[ 
\eta^{\alpha\beta} 
\partial_{\alpha} {\cal G} \partial_{\beta}{\cal G} - {{\cal G}'}^2
- \frac{z''}{z} {\cal G}^2\biggr].
\label{canscal}
\end{equation}
The results \cite{n1} is non trivial \cite{n1}.
These scalar normal modes are analogous to the ones we would obtain 
in the case of compact extra-dimensions \cite{mg}.

It should be appreciated that these equations are completely general
and do not depend on the specific background but only upon the general
form of the metric and of the action (\ref{ac}). In fact, in order
to derive Eqs. (\ref{ph})--(\ref{G}) and (\ref{def})--(\ref{z}) 
no specific background has been assumed, but only Eqs. (\ref{c1})--(\ref{c2}) 
which come directly from Eq. (\ref{ac}) and hold for any choice of the 
potential generating the scalar brane configuration.

The effective ``potentials'' appearing 
in the Schr\"odinger-like equations (\ref{ph}) and (\ref{G}) are dual 
with respect to $z\rightarrow 1/z$. This property can be used in order to 
discuss the properties of the massive spectrum \cite{n1}. Here 
it is sufficient to notice that the effective potentials can be related 
to their supersymmetric partner potentials \cite{n1} usually 
defined in the context of supersymmetric quantum mechanics \cite{susqm}. 
Under the duality transformation $z \rightarrow 1/z$ 
the superpotentials related to the equation for $\Phi$ and ${\cal G}$ 
go one into the other \cite{n1}. 

Let us now discuss the localization of the zero modes of the various 
fluctuations and enter the second step of the present discussion.
The lowest mass eigenstate of Eq. (\ref{mu}) 
is  $\mu(w) = {\mu_0} a^{3/2}(w)$. Hence, the normalization 
condition of the tensor zero mode implies 
\begin{equation}
|\mu_0|^2 \int_{-\infty}^{\infty} a^3 ~dw = 2 |\mu_0|^2\int_{0}^{\infty}a^3(w)
~ dw=1 .
\end{equation} 
where the assumed $w\rightarrow -w $ symmetry of the background geometry 
has been exploited. Using assumptions (i), (ii) and (iii) the tensor 
zero mode is then normalizable \cite{rs,rs2}. 

Let us now move to the analysis of vector fluctuations.
 Eq. (\ref{calV}) shows that the vector fluctuations are 
always massless and the corresponding zero mode is ${\cal V}(w) \sim 
{\cal V}_0 a^{-3/2} $. Consequently, the normalization 
condition will be 
\begin{equation}
2 |{\cal V}_0|^2 \int_{0}^{\infty} \frac{d w}{a^3(w)} =1,
\end{equation}
which cannot be satisfied if assumption (i), (ii) and (iii) hold. If $a^3(w)$ 
converges everywhere, $1/a^3(w)$ will not be convergent.
Therefore, if the four-dimensional Planck mass 
is finite the tensor modes of the geometry are normalizable and the vectors 
are not.

From Eq. (\ref{ph}) the lowest mass eigenstate 
of the metric fluctuation $\Phi$ corresponds to 
$\Phi(w)= \Phi_0 z^{-1}(w)$ and the 
related normalization condition reads 
\begin{equation} 
2|\Phi_0|^2 \int_{0}^{\infty} \frac{dw}{z^2(w)}=1.
\label{norm}
\end{equation}
The integrand appearing in Eq. (\ref{norm}) will now be shown 
to be non convergent at infinity if the geometry is regular. In fact 
 according to assumption (i)
\begin{eqnarray}
&& R = \frac{4}{a^2} ( 2 {\cal H}' + 3 {\cal H}^2),
\nonumber\\
&&R_{M N}R^{M N} = \frac{4}{a^4}( 4 {\cal H}^4 + 6 {\cal H}' {\cal H}^2
+ 5 {{\cal H}'}^2),
\nonumber\\
&& R_{M N A B}R^{M N A B} = \frac{8}{a^4}( 
2 {{\cal H}'}^2 - 5 {\cal H}^4),
\label{curv}
\end{eqnarray}
should be regular for any $w$ and, in particular, at infinity. The absence 
of poles in the curvature invariants guarantees the regularity of the 
five-dimensional geometry. 
Eq. (\ref{curv}) rules then out,  
warp factors decaying at infinity as $e^{-d w}$ or $e^{- d^2 w^2}$: these
profiles would lead to divergences in Eqs. (\ref{curv}) at infinity 
\footnote{In order to avoid confusions it should be stressed that 
exponential warp factors naturally appear in non-conformal coordinate systems
related to the one of Eq. (\ref{me}) as $a(w) dw = dy$.}.
Since $a(w)$ must converge at infinity,  $a(w) \sim w^{-\gamma}$
with $1/3\leq\gamma \leq 1$. Notice that $\gamma\geq 1/3$ 
comes  from the
 convergence (at infinity) of the integral of Eq. (\ref{pl}) \footnote{ 
In this sense the power $\gamma$ measures only the degree of convergence 
of a given integral.} and that $\gamma \leq 1$ is implied 
by Eqs. (\ref{curv}) since, at infinity, 
$R_{M N}R^{M N} \sim R_{M N A B} R^{M N A B } \sim 
w^{4 ( \gamma - 1)}$ should converge.
Using  Eq. (\ref{z}) and Eq. (\ref{c1}) the integrand of Eq. (\ref{norm})
can be written as 
\begin{equation}
\frac{1}{z^2} = \frac{{\cal H}^2}{a^3 {\varphi'}^2} = \frac{1}{6a^3}
\biggl(\frac{{\cal H}^2}{ {\cal H}^2 - {\cal H}'}\biggr).
\label{1/z}
\end{equation}
The behavior at infinity of Eq. (\ref{1/z}) can be now investigated assuming
the regularity of Eqs. (\ref{curv}), i.e. 
 $a(w) \sim w^{-\gamma}$ with $0< \gamma \leq 1$. In this limit 
\begin{equation}
\lim_{w\rightarrow \infty} \biggl(\frac{{\cal H}^2}{{\cal H}^2 
- {\cal H}'} \biggr)\sim \frac{\gamma^2}{\gamma^2 -\gamma} .
\label{lim1}
\end{equation}
and $1/z^2$ diverges {\em at least} as $a^{-3}$. In fact, if  $\gamma =1$, 
$1/z^2$ diverges even more as it can be argued from Eq. (\ref{lim1}) which 
has a further pole for $\gamma^2 = \gamma$. 
The  example given in Eqs. (\ref{s1})--(\ref{s2}) corresponds to 
a behavior at infinity given by $\gamma=1$.  
Direct calculations show that $1/z^2$ diverges, in this case,  as $w^5$.

Consequently, if the four-dimensional Planck mass is finite and if 
space-time is  regular the gauge-invariant (scalar) 
 zero mode is not normalizable and not localized on the brane.
For sake of completeness it should be mentioned that, for the lowest 
mass eigenvalue, 
there is a second (linearly independent) solution to Eq. (\ref{ph}) 
which is given by $ z^{-1}(w) \int^{w} z^2(x) ~dx$ which has poles at 
infinity and for $w\rightarrow 0$.
The poles appearing for $w\rightarrow 0$
will now be discussed since they are needed in order to prove that 
the zero modes of Eq. (\ref{G}) are not localized. As far as the poles 
 at infinity are concerned it is interesting to consider 
what happens to  
$ z^{-1}(w) \int^{w} z^2(x) ~dx$ in the case of the solution 
(\ref{s1})--(\ref{s2}). In this case, by direct use of Eqs. 
(\ref{s1})--(\ref{s2}) and (\ref{z}) we have that the second solution
diverges, at infinity, as $ (1 + b^2 w^2)^{1/4}\, (1 + 2 b^2 w^2)$. 

Noticing the duality connecting the effective 
potentials of Eqs. (\ref{ph}) and (\ref{G}) it can be
verified that the lowest mass eigenstate of Eq. (\ref{G}) is 
given by ${\cal G}(w)= {\cal G}_0 z(w)$.  
Provided the assumptions (i)--(iv) are satisfied,
it will now be demonstrated that the integral
\begin{equation}
2 |{\cal G}_{0}|^2 \int_{0}^{\infty} z^2 ~dw,
\end{equation}
is divergent not because of the behavior  at infinity
but because of the behavior of the solution close to the core 
of the wall, i.e. $w\rightarrow 0$. 
Bearing in mind Eq. (\ref{curv}),
assumption (i) and (ii) imply that $a(w)$ and $\varphi$ 
should be regular for any $w$. More specifically
close to the core of the wall $\varphi$ should go to zero 
and  $a(w)$ should go to a 
constant because of $w\rightarrow - w$ 
symmetry and the following regular expansions can be written
for small $w$
\begin{eqnarray}
&& a(w) \simeq a_0 - a_1\, w^{\beta} + ..., ~~~~~\beta>0,
\label{exp1}\\
&& \varphi(w) \simeq \varphi_1\,w^{\alpha} +..., ~~~~~\alpha >0,
\label{exp2}
\end{eqnarray}
for $w\rightarrow 0$. Inserting 
the expansion (\ref{exp1})--(\ref{exp2}) 
into Eq. (\ref{c1}) the relations can be obtained:
\begin{equation}
\beta = 2\alpha, \,\,\,\, \alpha^2 \varphi_1^2 = 6 \frac{a_1}{a_0} 
~\beta(\beta-1).
\label{cond}
\end{equation}
Inserting now Eqs. (\ref{exp1})--(\ref{exp2}) into Eq. (\ref{z}) 
and exploiting the first of Eqs. (\ref{cond}) we have
\begin{equation}
\lim_{w\rightarrow 0} z^2(w) \simeq w^{2 (\alpha - \beta)} = w^{-2\alpha}.
\,\,\,\alpha >0
\label{divv}
\end{equation}
Using Eqs. (\ref{exp1}) into 
Eqs. (\ref{curv}), $R_{A B} R^{A B} \sim 
R_{M N A B} R^{ M N A B} \sim w^{ 2 \beta -4}$, which implies $\beta \geq 2$ 
in order to have regular invariants for $w \rightarrow 0$.  
Since, from Eq. (\ref{cond}),  $\beta = 2\alpha$, in Eq. (\ref{divv})  
it must be $\alpha \geq 1$.
As in the case of Eq. (\ref{ph}) also eq. (\ref{G}) has a second 
(linearly independent) solution for the lowest mass eigenvalue, namely 
$z(w) \int^{w} dx~z^{-2}(x)$ which has poles at infinity.  
In fact, a direct check shows that, at infinity, this quantity
 goes ar $w^{\frac{3}{2}\gamma+ 1}$ where, as usual, 
$1/3\leq\gamma\leq 1$ for the convergence of the Planck mass and of the 
curvature invariants at infinity. 

We showed that the graviphoton and the graviscalars are delocalized 
under the assumptions (i)--(iv). The delocalization of a given 
zero mode may also be interpreted as a break-down of perturbation theory
since the lowest mass eigenstate is divergent for such a mode 
\footnote{ I thank G. Veneziano for stressing this point \cite{GV}.}. 
The interpretation of this phenomenon is that zero-modes whose 
wave functions are singular decouple from the four-dimensional 
effective theory. One can wonder if these divergences could be 
regularized in a gauge-invariant way. A standard way of implementing 
this regularization procedure is to include quadratic 
corrections to the Einstein-Hilbert term hoping that 
the divergences of the zero modes will disappear because of the 
different equations of the fluctutaions.
The same analysis performed here can be extended to the context of 
theories with higher derivatives \cite{n3}. In these theories, 
brane solutions can be indeed found both in five \cite{n3,z} 
and in six dimensions\cite{gm,hh}.
The  conclusions obtained are similar to the ones reported in the
present analysis.

\renewcommand{\theequation}{4.\arabic{equation}}
\setcounter{equation}{0}
\section{Localization of the modes of the geometry} 
As far as the localization properties 
of the various modes of the geometry are concerned, 
the convergence of the following integrals should 
be checked:
\begin{eqnarray}
&& I_{{\rm tens}}= \int_{0}^{+\infty} a^d(w) dw,
\label{first1}\\
&& I_{\rm vec} =
\int_{0}^{\infty} \frac{d w}{a^d(w)} ,
\label{first2}\\
&& I_{\Phi}= \int_{0}^{+\infty} \frac{dw}{z^2(w)},
\label{sec1}\\ 
&& I_{{\cal G}}=\int_{0}^{+\infty} z^{2}(w) dw,
\label{sec2}
\end{eqnarray}
where 
\begin{equation}
z(w) = \frac{a^{d/2} \varphi'}{{\cal H}},
\end{equation}
and with $d=3$ in the case of a four-dimensional Poincar\'e invariant 
world.
It has been demonstrated that under 
assumptions (i)--(iv), the scalar and vector fluctuations
 of the five-dimensional metric decouple from the wall. 
Eqs. (\ref{first1})--(\ref{sec2}) do not assume 
any {\em specific} background solution
but only the form (\ref{me}) of the metric 
together with the background equations. Similarly
the obtained results have been 
discussed and obtained in {\em general terms}.
General means that in order to assess the conclusions presented 
here no particular solution has been used. 
The analysis  of the fluctuations is independent 
on the specific coordinate system. The gauge-invariant method 
proposed in this context can certainly be applied to
other (related) contexts.

Heeding experimental tests \cite{exp}, 
the present results suggest that, under the 
assumptions (i)--(iv), no vector or scalar component of the Newtonian 
potential at short distances should be expected.

\Acknowledgements
The author wishes to thank M. E. Shaposhnikov 
for important discussions.


\begin{thebibliography}{99}
\bibitem{m1} V. A. Rubakov and M. E. Shaposhnikov, Phys. Lett. 
B {\bf 125}, 136
(1983).

\bibitem{m2} V. A. Rubakov and M. E. Shaposhnikov, Phys. Lett. B {\bf 125}, 139
(1983).

\bibitem{ak} K. Akama, in {\em Proceedings of the 
Symposium on Gauge Theory and Gravitation}, Nara, Japan, 
eds. K. Kikkawa, N. Nakanishi and H. Nariai, (Springer-Verlag, 1983),
[hep-th/0001113].

\bibitem{vis} M. Visser, Phys. Lett. B {\bf 159} (1985) 22.

\bibitem{misproc} M. Shaposhnikov, these proceedings. 

\bibitem{rs} L. Randall and R. Sundrum, Phys. 
Rev. Lett. {\bf 83} 3370 (1999). 

\bibitem{rs2}  L. Randall and R. Sundrum, Phys. 
Rev. Lett. {\bf 83} 4690 (1999). 

\bibitem{rub} V. A. Rubakov,  hep-ph/0104152.

\bibitem{bar} J. M. Bardeen, Phys. Rev. D {\bf 22}, 1882 (1980).

\bibitem{n1} M. Giovannini, Phys.Rev.D {\bf 64}, 064023 (2001). 

\bibitem{n2} M. Giovannini, hep-th/0106131. 

\bibitem{susqm} F. Cooper, A. Khare, and U. Sukhatme, Phys. Rep. 
{\bf 251}, 265 (1995).

\bibitem{kt1} A. Kehagias and K. Tamvakis, Phys.Lett. B {\bf 504}, 38 (2001).

\bibitem{kt2} A. Kehagias and K. Tamvakis, hep-th/0011006.

\bibitem{gremm1} M. Gremm, Phys. Lett. B {\bf 478}, 434 (2000).

\bibitem{gremm2} M. Gremm,  Phys. Rev. D {\bf 62}, 044017 (2000).

\bibitem{free} O. DeWolfe, D.Z. Freedman, S.S. Gubser, A. Karch, 
Phys.Rev.D {\bf 62}, 046008 (2000).  

\bibitem{free2} O. DeWolfe, D. Z. Freedman, hep-th/0002226. 

\bibitem{gms} M. Giovannini, H. Meyer, and M. Shaposhnikov, hep-th/0104118. 

\bibitem{gm} M. Giovannini, H. Meyer, hep-th/0108156 (Phys. Rev. D, in press).

\bibitem{kim} J. E. Kim,  B. Kyae, and H. M. Lee, hep-th/0110103. 

\bibitem{n3} M. Giovannini, Phys.Rev.D {\bf 64}, 124004 (2001). 

\bibitem{mg} M. Giovannini,  Phys. Rev. D {\bf 55}, 595 (1997).

\bibitem{GV} G. Veneziano, private communication.

\bibitem{z} O. Corradini and Z. Kakushadze, Phys. Lett. B {\bf 494}, 302
 (2000).
\bibitem{hh} M. Giovannini,  Phys.Rev.D {\bf 63}, 064011 (2001); 
Phys.Rev.D {\bf 63}, 085005 (2001).  

\bibitem{exp} C. D. Hoyle et al., Phys. Rev. Lett. {\bf 86} (2001) 1418.

\end{thebibliography}
\end{document}